 \documentclass[page-classic]{epl2}

\title{Static structure factor of a strongly correlated Fermi gas at large momenta}

\author{H. Hu\inst{1,2}\thanks{hhu@swin.edu.au}\and X.-J. Liu\inst{1} \and P. D. Drummond\inst{1}}

\institute{
\inst{1} ARC Centre of Excellence for Quantum-Atom Optics, Centre
for Atom Optics and Ultrafast Spectroscopy, Swinburne University of Technology, Melbourne 3122, Australia \\
\inst{2} Department of Physics, Renmin University of China, Beijing
100872, China}


\abstract{
We theoretically investigate  the static structure factor of an interacting
Fermi gas near the BEC-BCS crossover at large momenta. Due to
short-range two-body interactions, we \revision{show} that the structure
factor of unlike spin correlations $S_{\uparrow\downarrow}(q)$ falls off as
$1/q$ in a universal scaling region with large momentum $\hbar q$
and large scattering length. The scaling coefficient is determined
by the celebrated Tan's contact parameter, which links the short-range behavior
of many-body systems to their universal thermodynamic properties.
By implementing this \revision{structure-factor} Tan relation together with the random-phase approximation
and the virial expansion theory in various limiting cases, we show
how to calculate $S_{\uparrow\downarrow}(q)$ at zero and finite temperatures
for arbitrary interaction strengths, at momentum transfer higher than
the Fermi momentum. Our results provide a way to  experimentally confirm
\revision{one of the Tan relations} and to  accurately measure
the value of contact parameter.}

\begin{document}

\maketitle

Strongly interacting fermions are an important feature of many physical systems
in condensed matter physics, nuclear physics and astrophysics. The
recent achievements of trapping and cooling ultracold fermionic atoms
have made them significant in atomic physics as well \cite{rmp}.
With \revision{the} unprecedented
controllability of interactions and geometries \cite{rmp}, these atomic systems are
prototypes of many important theoretical models. 
The strength of the interactions is governed by a single (i.e., $s$-wave)
two-body scattering length, and can be tuned by means of Feshbach
resonances, allowing for a systematic exploration of the crossover from
a weakly coupling BCS superfluid to a Bose-Einstein condensate (BEC)
of tightly-bound molecules. Ultracold atoms therefore provide
a new platform for investigating the intriguing many-body properties of
fermions. A rich variety of theoretical predictions has emerged, many
of which await verification. Theoretical challenges arise especially when the
scattering length $a$ is comparable to or larger than the inter-particle
distance $l$ \cite{thomas}. In this strongly interacting
regime, the system has universal scaling properties that depend on
$l$ only \cite{ho,ournatphys}. These universal properties, however,
cannot fully be understood using conventional perturbation methods,
which are reliable only for $a\ll l$ \cite{weakcouplingtheory,ournsr}.
Ab-initio quantum Monte Carlo (QMC) simulations are very helpful \cite{qmceos,qmcssf},
but suffer from the Fermi sign problem in many cases. The challenging
nature of strong interactions therefore makes \emph{exact} results very valuable.

In this Letter, we \revision{discuss theoretically} an exact relation for the large-momentum
behavior of the spin-antiparallel static structure factor $S_{\uparrow\downarrow}(q)$,
showing that it has a simple universal power-law ($1/q$) tail. \revision{This power-law 
behavior was recently confirmed at Swinburne University, using $^{6}$Li 
atoms near a Feshbach resonance at nearly zero temperature \cite{ourtan}. 
Here, we present the theoretical details}.

\revision{The relation we consider, hereafter referred to as the structure-factor Tan relation, belongs} to 
the family of exact relations obtained by Tan in 2005 \cite{tan}, which link the short-range, large-momentum,
and high-frequency asymptotic behavior of many-body systems to their
thermodynamic properties \cite{tan,rfsumrule,werner,taylor,son}. For instance,
the momentum distribution and rf spectrum fall off as $q^{-4}$ and
$\omega^{-5/2}$, respectively. All the Tan relations are related
to each other by a single\emph{ }coefficient ${\cal I}$, referred
to as the integrated contact density or contact. The
contact measures the probability of two fermions with unlike spins
being close together \cite{braatenplatter}. It also links this short-range behaviour to thermodynamics
via the adiabatic relation, $dE/d(-1/a)=\hbar^{2}\mathcal{I}/(4\pi m)$,
which gives the change in the total energy $E$ due to adiabatic changes
in the scattering length. The fundamental importance of the Tan relations is 
due to their wide applicability: to zero or finite
temperature, superfluid or normal phase, homogeneous or
trapped, few-body or many-body systems.

\revision{The structure-factor} Tan relation for $S_{\uparrow\downarrow}(q)$ follows directly from the short-range
behavior of the pair correlation function $n_{\uparrow\downarrow}^{(2)}({\bf r})\equiv\int d{\bf R}\left\langle \hat{n}_{\uparrow}\left({\bf R}-\mathbf{r}/2\right)\hat{n}_{\downarrow}\left({\bf R}+\mathbf{r}/2\right)\right\rangle $,
which diverges as 
\begin{equation}
n_{\uparrow\downarrow}^{(2)}({\bf r}\rightarrow0)\simeq\frac{{\cal I}}{16\pi^{2}}\left(\frac{1}{r^{2}}-\frac{2}{ar}\right).\end{equation}
A Fourier transformation of $n_{\uparrow\downarrow}^{(2)}({\bf r}\rightarrow0)$
then leads to
\begin{equation}
S_{\uparrow\downarrow}\left(q\gg k_{F}\right)\simeq\frac{{\cal I}}{4Nq}\left[1-\frac{4}{\pi aq}\right]\equiv\frac{{\cal I}}{Nk_{F}}t\left(q\right),\label{TanSSF}\end{equation}
where $k_{F}$ is the Fermi wave-vector and $N$ is the total number
of atoms. On the right hand side of the above equation, we have defined
$t(q)\equiv[k_{F}/(4q)][1-4/(\pi aq)]$. Eq. (\ref{TanSSF})
holds in a scaling region of sufficiently large $q$ near the
unitarity limit ($a\rightarrow\pm\infty$) so that the next-order
correction in the bracket ($\propto1/(aq)$) is small compared to
the leading term of $1$. \revision{However, the momentum $q$ should always be much smaller 
than the inverse of the range of the interaction potential, so that the pair correlation function has a  
$1/r^2$ behavior. Eqs. (1) and (2) were discussed briefly in Ref. \cite{ourtan}. }

The power-law tail of $1/q$ \revision{in the structure-factor Tan relation} is more amenable
for experimental investigation than the $q^{-4}$ or $\omega^{-5/2}$
tail \revision{in the Tan relations for the momentum distribution or using rf spectroscopy}. The fast decay due to the higher power law
in these latter two cases imposes very stringent signal-to-noise requirements for studying a given range of momenta or frequency.
Experimentally, the static structure factor $S(q)=S_{\uparrow\uparrow}(q)+S_{\uparrow\downarrow}(q)$
can be readily measured using two-photon Bragg spectroscopy on a
balanced two-component atomic Fermi gas near a Feshbach resonance\cite{swinexpt}.
In the large-$q$ limit, to an good approximation $S_{\uparrow\uparrow}(q)\simeq1$.
One thus can directly determine  the spin-antiparallel structure factor
$S_{\uparrow\downarrow}(q)=S(q)-1$ and verify a simple $1/q$ asymptotic
behavior, \revision{as shown in Ref. \cite{ourtan}}.

The purpose of this Letter is to quantify the size of the universal scaling region
where the \revision{structure-factor} Tan relation (\ref{TanSSF}) is valid. For this goal, it
is desirable to calculate $S_{\uparrow\downarrow}(q)$ at large momentum
(i.e., $q>k_{F}$) at \emph{arbitrary} interaction strength and
temperature. This is a challenging theoretical problem.
So far, the static structure factor of strongly interacting fermions has only
been studied using QMC simulations at zero temperature
in a homogeneous medium \cite{qmcssf}.

In the present work, we first revisit the zero-temperature problem by
extending the scaling result Eq. (\ref{TanSSF}) to the weak coupling
BEC and BCS limits with $\left|k_{F}a\right|\ll1$. In the BEC regime,
$S_{\uparrow\downarrow}(q)$ is mainly caused by ground
state molecules. In the opposite BCS regime, we use a random-phase
linear response theory to obtain first the dynamic and subsequently the static
structure factor. These two limits are then interpolated smoothly
to the unitarity limit. In this procedure, the zero-temperature
contact ${\cal I}$ is determined from a reliable theoretical prediction
of the ground state energy along the BEC-BCS crossover, via the adiabatic
Tan relation. As an alternative, at finite temperatures close to the Fermi temperature
($T\sim T_{F}$), we develop a quantum virial expansion theory and
calculate  the structure factor quantitatively at high temperature. In this high temperature
regime, \revision{the structure-factor} Tan relation (\ref{TanSSF}) can be analytically demonstrated,
and the temperature dependence of the contact is extracted. The interpolation
strategy adopted at zero temperature can also be strictly examined.

\textit{Weak coupling limit} --- We start by considering the weak
coupling BEC limit with a small and positive scattering length, $1/(k_{F}a)\rightarrow+\infty$.
Here the system is a dilute gas of weakly interacting bound molecules,
consisting of two atoms with opposite spins, with a relative wave
function $\Psi_{0}\left(r\right)=\exp[-r/a]/(\sqrt{2\pi a}r)$. Neglecting
the correlation among molecules at $k_{F}r<1$, the pair correlation
function is given by $n_{\uparrow\downarrow}^{(2)}({\bf r})\simeq\Psi_{0}^{2}\left(r\right)$,
leading to \cite{qmcssf},

\begin{equation}
\left[S_{\uparrow\downarrow}\left(q > k_{F}\right)\right]_{BEC}\simeq\frac{2}{qa}\arctan(\frac{qa}{2}),\label{SSFBEC}\end{equation}
which is essentially geometry and temperature independent. For $qa\gg1$,
the first two terms in the Taylor expansion of the expression agrees
exactly with Eq. (\ref{TanSSF}), since ${\cal I}_{BEC}/(Nk_{F})\simeq 4\pi/(k_{F}a)$
\cite{tan}.

In the opposite BCS limit of $1/(k_{F}a)\rightarrow-\infty$, we follow
a mean-field picture and treat the system as a gas of quasiparticles
subjected to a (dynamical) mean-field Hamiltonian \cite{textbook},
${\cal H}={\cal H}_{0}+(4\pi\hbar^{2}a/m)(\delta n_{\uparrow}+\delta n_{\downarrow})$,
where ${\cal H}_{0}$ is the free particle Hamiltonian and $\delta n_{\sigma}\equiv\delta n_{\sigma}({\bf r},t)$
are the space- and time-dependent density fluctuations with respect
to equilibrium. Here, we consider a normal state since the result
differs by an exponentially small amount from that of the BCS superfluid state.
The dynamic structure factor $S_{\sigma\sigma^{\prime}}(q,\omega;T)=-\mathop{\rm Im}\chi_{\sigma\sigma^{\prime}}(q,\omega;T)/[\pi(1-e^{-\hbar\omega/k_{B}T})]$
may be calculated via the density response functions within a (random-phase)
linear response theory \cite{textbook}, which are given by, $\chi_{\sigma\sigma^{\prime}}=\chi_{\sigma\sigma^{\prime}}^{0}+\sum_{\sigma^{^{\prime\prime}}\sigma^{^{\prime\prime\prime}}}\chi_{\sigma\sigma^{^{\prime\prime}}}^{0}(4\pi\hbar^{2}a/m)\chi_{\sigma^{^{\prime\prime\prime}}\sigma^{\prime}}$.

The free response function $\chi_{\sigma\sigma^{\prime}}^{0}$ can
be written in a simple form in terms of a dimensionless function $g(\tilde{q},\lambda;\tilde{T})$:
$\chi_{\sigma\sigma^{\prime}}^{0}(q,\omega;T)\equiv-[mk_{F}/(2\pi^{2})]g(\tilde{q},\lambda;\tilde{T})\delta_{\sigma\sigma^{\prime}}$,
with $\tilde{q}\equiv q/k_{F},$ $\lambda\equiv m\omega/(\hbar qk_{F}),$
and $\tilde{T}\equiv T/T_{F}$. The function $g(\tilde{q},\lambda;\tilde{T})$
can be found in standard textbooks \cite{textbook}. To  leading
order in $k_{F}a$, it is straightforward to show that for a \emph{homogeneous}
Fermi gas, the static structure factor $S_{\uparrow\downarrow}\left(q\right)\equiv(2/N)\int d\omega S_{\uparrow\downarrow}(q,\omega;T)$
is given by 
\begin{equation}
\left[S_{\uparrow\downarrow}\left(q\right)\right]_{BCS}\simeq-\frac{12k_{F}a}{\pi^{2}}\tilde{q}\int\limits _{-\infty}^{\infty}d\lambda\frac{\mathop{\rm Re}g(\tilde{q},\lambda;\tilde{T})\mathop{\rm Im}g(\tilde{q},\lambda;\tilde{T})}{1-\exp[-2\tilde{q}\lambda/\tilde{T}]}.\label{SSFBCS}\end{equation}

This weak-coupling equation holds for arbitrary transferred momentum.
We may rewrite it into the form, $[S_{\uparrow\downarrow}(q)]_{BCS}\simeq{\cal I}_{BCS}/(Nk_{F})[-4/(\pi qa)]f(\tilde{q};\tilde{T})/(4q/k_{F})$,
where at $T=0$ we have defined $f(\tilde{q};\tilde{T}=0)\equiv(9/\pi)\tilde{q}^{3}\int_{0}^{\infty}d\lambda\mathop{\rm Re}g(\tilde{q},\lambda;0)\mathop{\rm Im}g(\tilde{q},\lambda;0)$
and have confirmed that ${\cal I}_{BCS}/(Nk_{F})\simeq4(k_{F}a)^{2}/3$
\cite{tan}. At large momentum, asymptotically, $f(\tilde{q}\gg1;\tilde{T}=0)=1+2/(5\tilde{q}^{2})+O(\tilde{q}^{-4})$.
Thus, at the leading order of interaction strengths ($\sim k_{F}a$),
the structure factor in the BCS limit is in agreement with the Tan
relation (\ref{TanSSF}). We note, however, that the dominant $1/q$
tail in the strongly interacting regime (i.e., the factor 1 in the
square bracket in Eq. (\ref{TanSSF})) is lost. The large momentum tail  comes
from  pairing effects and therefore cannot be captured by the random-phase
approximation for linear response.

\begin{figure}
\onefigure[clip,width=0.70\textwidth]{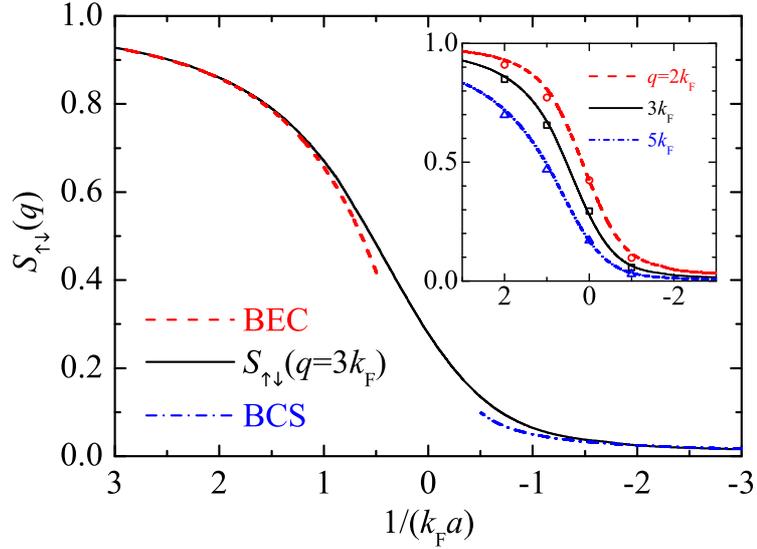} 
\caption{(Color on-line) Zero temperature ($T=0$) spin-antiparallel static
structure factor of a homogeneous Fermi gas along the BEC-BCS crossover
at $q=3k_{F}$, calculated by using the interpolation strategy as
described in the text (Eq. \ref{TanSSFIP}). The red dashed and blue
dot-dashed line show the asymptotic behavior in the weak-coupling
BEC (Eq. \ref{SSFBEC}) and BCS (Eq. \ref{SSFBCS}) regime, respectively.
The inset compares our interpolation results (lines) with the QMC simulations (symbols) \cite{qmcssf}. 
The excellent agreement at large momentum justifies our interpolation idea.}
\label{fig1} 
\end{figure}

\textit{Structure factor from an interpolation strategy}. --- The excellent
agreement between the large-$q$ expansion of the weak coupling results,
Eqs. (\ref{SSFBEC}) and (\ref{SSFBCS}), and the Tan relation (\ref{TanSSF})
suggests that we may use the following extrapolation for the calculation
of the structure factor at $q>k_{F}$ and at arbitrary interaction
strengths:

\begin{equation}
S_{\uparrow\downarrow}\simeq\left(\frac{{\cal I}}{Nk_{F}}\right)\times\left\{ \begin{array}{ll}
k_{F}/\left(4q\right)\left[1-4f(q/k_{F};T/T_{F})/\left(\pi aq\right)\right], & a<0;\\
k_{F}/(2\pi q)\arctan(qa/2), & a>0.\end{array}\right.\label{TanSSFIP}\end{equation}
In doing so, we remove the severe requirements of $q\gg k_{F}$ and
$q\gg1/a$ for validating the Tan relation (\ref{TanSSF}). In Fig.
1, we show the resulting static structure factor $S_{\uparrow\downarrow}\left(q=3k_{F}\right)$
for a homogeneous Fermi gas along the BEC-BCS crossover. We have determined
the contact ${\cal I}$ by applying the adiabatic relation to the
ground state energy calculated from a perturbative Gaussian pair fluctuation
theory \cite{ournsr}. In the figure, the weak coupling results in
Eqs. (\ref{SSFBEC}) and (\ref{SSFBCS}) are also reported respectively
by the dashed and dot-dashed lines. By applying a local density approximation,
we may also calculate the zero temperature structure factor of a Fermi
gas in a harmonic trap with frequency $\omega_{0}$ and length $a_{ho}=\sqrt{\hbar/m\omega_{0}}$.
Using the ideal Fermi temperature and vector at the trap center, $k_{B}T_{F}=\left(3N\right)^{1/3}\hbar\omega_{0}$
and $k_{F}=(24N)^{1/6}/a_{ho}$, we find that the trapped structure
factor differs only slightly with the uniform one (see the black line
in Fig. 2).

The accuracy of the interpolation  may be justified by comparing
our results to  QMC  simulations \cite{qmcssf},
as shown in the inset for $q=2k_{F}$, $3k_{F}$, and $5k_{F}$. We
find  a large deviation occurring at $q < k_{F}$ on the
BEC side (not shown in the inset), where one has to take into account
the correlations between different bound molecules. However, this is irrelevant to our purpose 
of examining large momentum correlations. As we shall see, a
further justification of our interpolation scheme can be obtained
at finite temperature, where a quantitative calculation of the contact
and structure factor is feasible by using a quantum virial expansion.
In this temperature regime, a random-phase linear response theory
predicts for a uniform gas, 
\begin{equation}
f(\tilde{q};\tilde{T})=[\tilde{q}/(\pi\sqrt{\tilde{T}})]\int_{-\infty}^{+\infty} u(x)\exp[-(x-\tilde{q}/\sqrt{\tilde{T}})^{2}] dx,
\end{equation}
where the function $u(x)\equiv\int_{0}^{\infty}dt[e^{-(x-t)^{2}}-e^{-(x+t)^{2}}]/t$.
We find that $f(\tilde{q}\gg1;\tilde{T})=1+\tilde{T}/(\tilde{q}^{2})+O(\tilde{q}^{-4})$.

\textit{Quantum virial expansion at finite temperatures}. --- We now
calculate the structure factor at high temperatures for a \emph{trapped}
Fermi gas, by extending a previous virial expansion theory \cite{ourve,vedsf,veakw}.
At high temperatures, the fugacity $z\equiv\exp(\mu/k_{B}T)\ll1$
is a small controllable parameter, even in the strongly interacting
limit. We may expand the thermodynamic potential $\Omega=-k_{B}T\ln{\cal Z}$,
where ${\cal Z}={\rm Tr}\exp[-({\cal H}-\mu{\cal N})/k_{B}T]$, in
a series of powers of fugacity. By defining partition functions of
clusters in the $n$-particle subspace, $Q_{n}={\rm Tr}_{n}[\exp(-{\cal H}_{n}/k_{B}T)]$,
we find that $\Omega/k_{B}T=-[zQ_{1}+z^{2}\left(Q_{2}-Q_{1}^{2}/2\right)+\cdots]$.
The pair correlation function $n_{\uparrow\downarrow}^{(2)}({\bf R=}({\bf r}_{1}+{\bf r}_{2})/2,{\bf r=r}_{1}-{\bf r}_{2})\equiv$
$\left\langle \hat{n}_{\uparrow}\left({\bf r}_{1}\right)\hat{n}_{\downarrow}\left({\bf r}_{2}\right)\right\rangle $
can be calculated by taking a functional derivative of the thermodynamic
potential with respect to the generating elements of a source term
$\int d{\bf r}_{1}d{\bf r}_{2}\vartheta_{\uparrow}({\bf r}_{1})\vartheta_{\downarrow}({\bf r}_{2})\hat{n}_{\uparrow}\left({\bf r}_{1}\right)\hat{n}_{\downarrow}\left({\bf r}_{2}\right)$,
i.e., $n_{\uparrow\downarrow}^{(2)}({\bf r}_{1},{\bf r}_{2})=\left.\delta^{2}\Omega(\vartheta_{\uparrow},\vartheta_{\downarrow})/(\delta\vartheta_{\uparrow}\delta\vartheta_{\downarrow})\right|_{\vartheta=0}$.
Tthe virial expansion
form of the thermodynamic potential is particularly suitable for taking
a derivative of  to $Q_{n}$, and hence for applying the Hellmann--Feynman theorem. 

To leading order, it is readily seen that $n_{\uparrow\downarrow}^{(2)}({\bf R},{\bf r})=z^{2}\sum_{P}\exp(-E_{P}/k_{B}T)\left|\Psi_{P}({\bf R},{\bf r})\right|^{2}$,
where the summation is over all the two-body states $\Psi_{P}({\bf R},{\bf r})$
of a pair made of atoms with opposite spins. In a harmonic trap, the
center-of-mass and relative motion parts of the wavefunction are separable.
By taking the Fourier transform of the relative coordinate (${\bf r}$)
and performing a spatial integral over the center-of-mass part (${\bf R}$),
we finally arrive at\begin{equation}
S_{\uparrow\downarrow}\left(q\right)=\frac{2z^{2}}{N\left[e^{+\tilde{\beta}/2}-e^{-\tilde{\beta}/2}\right]^{3}}\int_{0}^{\infty}dr\frac{\sin qr}{qr}\left[\sum_{n}\exp(-\frac{E_{rel,n}}{k_{B}T})\psi_{rel,n}^{2}(r)-\{Non.\}\right],\label{SSFVE}\end{equation}
where $\tilde{\beta}=\hbar\omega_{0}/k_{B}T$, and $\{Non.\}$ represents the same sum as the first term in the bracket, but for a non-interacting system.
The relative energies $E_{rel,n}$
and radial wavefunctions $\psi_{rel,n}(r)$ are known
\cite{ourve}: $E_{rel,n}=(2\nu_{n}+3/2)\hbar\omega_0$, where $\nu_{n}$
is given by $\sqrt{2}\Gamma(-\nu_{n})/\Gamma(-\nu_{n}-1/2)=a_{ho}/a$,
and $\psi_{rel,n}(r)=A_{n}(r/a_{ho})\exp[-r^{2}/(4a_{ho}^{2})]\Gamma(-\nu_{n})U(-\nu_{n},3/2,r^{2}/(2a_{ho}^{2}))$,
where $A_{n}$ is the normalization factor, and $\Gamma$ and $U$
are respectively the Gamma and confluent hypergeometric functions.

The accuracy of the virial expansion (\ref{SSFVE}) may be examined
by applying the same expansion to the equation of state \cite{unitaritycmp},
which turns out to be quantitatively reliable at $T\sim T_{F}$. It
is straightforward to improve the accuracy by including contributions
from $Q_{3}$, $Q_{4}$, and so on. However, considerable insight
can already be obtained from the leading two-body contribution shown
in Eq. (\ref{SSFVE}). At short distance, $\psi_{rel,n}^{2}(r\rightarrow0)\simeq2\pi A_{n}^{2}[1-2r/a-(2\nu_{n}+3/2)(r^{2}/a_{ho}^{2})+r^{2}/a^{2}+(4/3)(2\nu_{n}+3/2)r^{3}/(a_{ho}^{2}a)]$.
Therefore, for  large  momentum transfer $q$, we find from Eq. (\ref{SSFVE})
that,

\begin{equation}
S_{\uparrow\downarrow}\left(q\right)=\frac{{\cal I}_{VE}}{4Nq}\left[1-\frac{4}{\pi aq}+\frac{16}{3\pi aq^{3}a^2_{ho}}\frac{d\ln B}{d\tilde{\beta}}\right],\label{TanSSFVE}\end{equation}
where $B(\tilde{\beta})\equiv\sum_{n}\exp[-\tilde{\beta}(2\nu_{n}+3/2)]2\pi A_{n}^{2}$
and \begin{equation}
{\cal I}_{VE}=4\pi z^{2}\frac{1}{\left[e^{+\tilde{\beta}/2}-e^{-\tilde{\beta}/2}\right]^{3}}B(\tilde{\beta})\end{equation}
 is the contact predicted by the virial expansion. We thus explicitly
confirm \revision{the structure-factor} Tan relation (\ref{TanSSF}) within second order virial expansion
theory.

In the BEC limit, the two-body bound state $\psi_{rel,n}(r)\simeq\Psi_{0}\left(r\right)=\exp[-r/a]/(\sqrt{2\pi a}r)$
dominates the summation in Eq. (\ref{SSFVE}), with  energy $E_{rel,n}\simeq-\epsilon_{B}\equiv-\hbar^{2}/ma^{2}$.
The virial expansion of the number equation gives $[z^{2}/(e^{+\tilde{\beta}/2}-e^{-\tilde{\beta}/2})^{3}]\exp(\epsilon_{B}/k_{B}T)\simeq N/2$ \revision{at high temperatures}.
By inserting these results into Eq. (\ref{SSFVE}), we correctly recover 
the structure factor in the weak-coupling BEC limit, Eq. (\ref{SSFBEC}).  
At unitarity, the normalization factor $A_{n}$ is analytically known
and we find that $B(\tilde{\beta})=\sqrt{2/\pi}/(e^{+\tilde{\beta}}-e^{-\tilde{\beta}})^{1/2}\simeq\sqrt{k_{B}T/(\pi\hbar\omega_{0})}$.
Thus, at \revision{high temperatures} $T\geq T_{F}$ where $z\simeq(N/2)/(e^{+\tilde{\beta}/2}-e^{-\tilde{\beta}/2})^{3}\simeq(1/6)(T/T_{F})^{-3}$,
the contact is given by ${\cal I}_{VE,\infty}/(Nk_{F})\simeq(\sqrt{2\pi}/6)(T/T_{F})^{-5/2}$, and therefore  rapidly
decreases with temperature. The structure factor is simply $S_{\uparrow\downarrow}(q)={\cal I}_{VE,\infty}/(4Nq)$$[1-4/(\pi aq)]$.
Finally, in the BCS limit, we may also work out analytically that
$B(\tilde{\beta})=\sqrt{2/\pi}a^{2}/(e^{+\tilde{\beta}}-e^{-\tilde{\beta}})^{3/2}$.
This leads to ${\cal I}_{VE,BCS}/(Nk_{F})=[\sqrt{2\pi}\left(T/T_{F}\right)^{-3/2}/24](k_{F}a)^{2}$,
agreeing exactly with our prediction from  random-phase linear
response theory \revision{at high temperatures}. The resulting structure factor is given by \begin{equation}
S_{\uparrow\downarrow}\left(q\right)=\frac{{\cal I}_{VE,BCS}}{4Nq}\left[1-\frac{4}{\pi aq}-\frac{4Tk_{F}^{2}}{\pi T_{F}aq^{3}}\right].\label{eq:}\end{equation}

\begin{figure}
\onefigure[clip,width=0.70\textwidth]{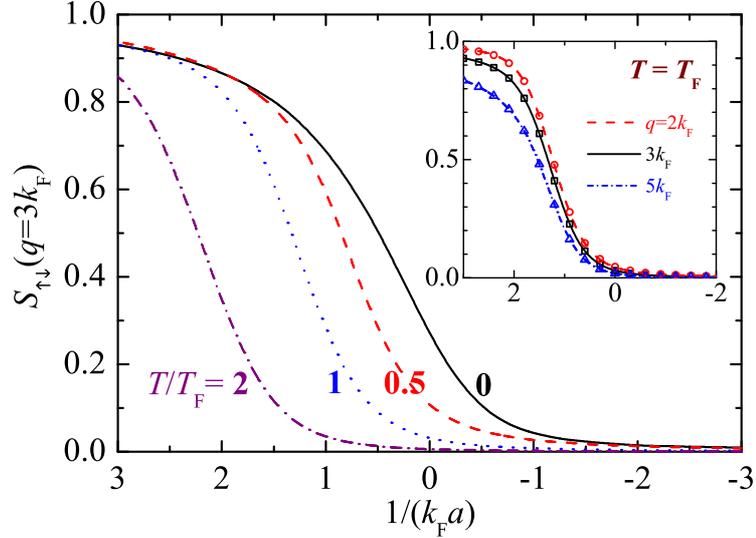} 
\caption{(Color on-line) Temperature dependence of spin-antiparallel static
structure factor in a harmonic trap. The zero temperature result is
obtained by applying a local density approximation to the homogeneous
structure factor. The inset shows a comparison of the leading virial
expansion predictions (symbols, Eq. \ref{SSFVE}) with the interpolation
strategy (lines, Eq. \ref{TanSSFIP}) where $f(\tilde{q};\tilde{T})$ is
averaged over the trap. This excellent agreement therefore justifies
the interpolation strategy at finite temperatures.}
\label{fig2} 
\end{figure}

In Fig. 2, we present the numerical results of the leading virial
expansion for the static structure factor ($q=3k_{F}$) of a trapped
Fermi gas. The fugacity $z$ is determined by the number equation,
$N=-\partial\Omega/\partial\mu$, expanding up to the second order
virial coefficient. We show also the zero temperature result obtained
from the local density approximation. The temperature dependence of
the structure factor is evident, particularly on the BCS side. In
the inset, we check explicitly the interpolation strategy with $f(\tilde{q};\tilde{T})$.
For $q>2k_{F}$, the extrapolation results (symbols) are indistinguishable
to the numerical virial expansion calculations (lines).

\begin{figure}
\onefigure[clip,width=0.80\textwidth]{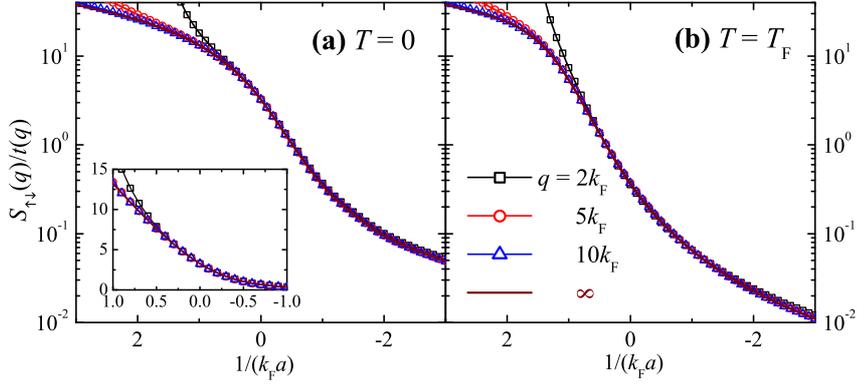} 
\caption{(Color on-line) Scaled spin-antiparallel static structure factor at
$T=0$ (a) and $T=T_{{\bf F}}$ (b) for a trapped Fermi gas. Inset
in (a): a blow-up of the $T=0$ scaled structure factor at the BEC-BCS
crossover regime. In the universal region with large momenta and strong
interactions, the scaled spin-antiparallel structure factor converges
to the dimensionless contact $\mathcal{I}/(Nk_{F})$.}
\label{fig3} 
\end{figure}

\textit{Universal scaling region of our Tan relation}. --- We are
now ready to determine the scaling region in which we expect that
the scaled structure factor $S_{\uparrow\downarrow}\left(q\right)/t\left(q\right)$
becomes momentum independent and converges to the contact of the system.
Fig. 3 shows the scaled structure factors of a trapped Fermi gas at
different transferred momenta and at low and high temperatures. On
the BCS side, the scaled structure factor is rather insensitive to
the varying momentum. We observe that the scaling limit can be easily
reached at the BEC-BCS crossover regime ($\left|1/k_{F}a\right| < 0.5$),
at a relatively small momentum $q=2k_{F}$. However, to access the
scaling limit in a broader region of interaction strengths (i.e., $1/k_{F}a < 2$),
a larger momentum ($q=5k_{F}$) is necessary.

\textit{Conclusion}. --- Based on \revision{the structure-factor} Tan relation \revision{Eq. (2)}, we present a
systematic study of the static structure factor of an interacting
Fermi gas near the BEC-BCS crossover at a transferred momentum $q>k_{F}$.
The scaling region of the \revision{structure-factor} Tan relation is clarified. These predictions
can be readily checked in future Bragg experiments at crossover, either near the ground state or at 
finite temperatures.

\acknowledgments
We thank C. J. Vale, E. D. Kuhnle, M. Mark, P. Dyke, and P. Hannaford for fruitful discussions. This
work was supported in part by the Australian Research Council (ARC)
Centre of Excellence for Quantum-Atom Optics, ARC Discovery Project
No. DP0984522 and No. DP0984637, NSFC Grant No. NSFC-10774190, and NFRPC
Grant No. 2006CB921404 and No. 2006CB921306.


\begin{thebibliography}{21}
  
\bibitem{rmp} 
\Name{Giorgini S., Pitaevskii L. P. \and Stringari S.}
\REVIEW{Rev. Mod. Phys.}{80}{2008}{1215}.

\bibitem{thomas} 
\Name{O'Hara K. M., Hemmer S. L., Gehm M. E., Granade S. R. \and Thomas J. E.}
\REVIEW{Science}{298}{2002}{2179}.

\bibitem{ho} 
\Name{Ho T.-L.}
\REVIEW{Phys. Rev. Lett.}{92}{2004}{090402}.

\bibitem{ournatphys} 
\Name{Hu H., Drummond P. D. \and Liu X.-J.}
\REVIEW{Nature Phys.}{3}{2007}{469}.

\bibitem{weakcouplingtheory} 
\Name{Nozi�res P. \and Schmitt-Rink, S.}
\REVIEW{J. Low Temp. Phys.}{59}{1985}{195};
\Name{Ohashi Y. \and Griffin A.}
\REVIEW{Phys. Rev. Lett.}{89}{2002}{130402};
\Name{Liu X.-J. \and Hu H.}
\REVIEW{Europhys. Lett.}{75}{2006}{364}.

\bibitem{ournsr} 
\Name{Hu H., Liu X.-J. \and Drummond P. D.}
\REVIEW{Europhys. Lett.}{74}{2006}{574}.

\bibitem{qmceos} 
\Name{Astrakharchik G. E., Boronat J., Casulleras J. \and Giorgini S.}
\REVIEW{Phys. Rev. Lett.}{93}{2004}{200404}.

\bibitem{qmcssf} 
\Name{Combescot R., Giorgini S. \and Stringari S.},
\REVIEW{Europhys. Lett.}{75}{2006}{695}.

\bibitem{ourtan}
\Name{Kuhnle E. D., Hu H., Liu X.-J., Dyke P., Mark M., Drummond P. D., Hannaford P. \and Vale C. J.}
\REVIEW{arXiv:1001.3200v2} {}{2010}{}.

\bibitem{tan} 
\Name{Tan S.}
\REVIEW{Ann. Phys.}{323}{2008}{2952};
\REVIEW{Ann. Phys.}{323}{2008}{2971};
\REVIEW{Ann. Phys.}{323}{2008}{2987}.

\bibitem{rfsumrule} 
\Name{Punk M. \and Zwerger W.}
\REVIEW{Phys. Rev. Lett.}{99}{2007}{170404};
\Name{Baym G., Pethick C. J., Yu Z. \and Zwierlein M. W.}
\REVIEW{Phys. Rev. Lett.}{99}{2007}{190407}.

\bibitem{werner} 
\Name{Werner F., Tarruell L. \and Castin Y.}
\REVIEW{Eur. Phys. J. B}{68}{2009}{401}.

\bibitem{taylor}
\Name{Taylor E. \and Randeria M.}
\REVIEW{Phys. Rev. A}{81}{2010}{053610}.

\bibitem{son}
\Name{Son D. T. \and Thompson E. G.}
\REVIEW{arXiv:1002.0922v1}{}{2010}{}.

\bibitem{braatenplatter} 
\Name{Braaten E. \and Platter L.}
\REVIEW{Phys. Rev. Lett.}{100}{2008}{205301}.

\bibitem{swinexpt} 
\Name{Veeravalli G., Kuhnle E. D., Dyke P. \and Vale C. J.}
\REVIEW{Phys. Rev. Lett.}{101}{2008}{250403}.

\bibitem{textbook} 
\Name{Fetter A. L. \and Walecka J. D.}
\Book{Quantum Theory of Many-Particle Systems}
\Publ{McGraw-Hill, New York}
\Year{1971}.

\bibitem{ourve}
\Name{Liu X.-J., Hu H. \and Drummond P. D.}
\REVIEW{Phys. Rev. Lett.}{102}{2009}{160401}.

\bibitem{vedsf}
\Name{Hu H., Liu X.-J. \and Drummond P. D.}
\REVIEW{Phys. Rev. A}{81}{2010}{033630}.

\bibitem{veakw}
\Name{Hu H., Liu X.-J. \and Drummond P. D.}
\REVIEW{Phys. Rev. Lett.}{104}{2010}{240407}.

\bibitem{unitaritycmp}
\Name{Hu H., Liu X.-J. \and Drummond P. D.}
\REVIEW{arXiv:1001.2085v1}{}{2010}{}; to be published in \textit{New Journal of Physics}. 

\end{thebibliography}
\end{document}